\documentclass{article}

\usepackage{lingmacros, fullpage, setspace, amsmath}     

\usepackage{arxiv}

\usepackage[utf8]{inputenc} 
\usepackage[T1]{fontenc}    
\usepackage{hyperref}       
\usepackage{url}            
\usepackage{booktabs}       
\usepackage{amsfonts}       
\usepackage{nicefrac}       
\usepackage{microtype}      
\usepackage{lipsum}		
\usepackage{graphicx}
\usepackage{natbib}
\usepackage{doi}
\usepackage{hyperref}       
\usepackage{booktabs,tabularx,threeparttable}

\newcommand{\bs}{\bigskip}

\usepackage{xcolor}
\definecolor{color5}{HTML}{006795}
\hypersetup{
  colorlinks   = true, 
  urlcolor     = blue, 
  linkcolor    = color5, 
  citecolor   = color5 
}
\title{The Black Tuesday Attack: \\ how to crash the stock market with adversarial examples to financial forecasting models}

\author{Thomas Hofweber\thanks{Corresponding author: hofweber@unc.edu} \/ \thanks{University of North Carolina at Chapel Hill}  \quad
Jefrey Bergl\footnotemark[1] \quad  Ian Reyes\footnotemark[1] \quad Amir Sadovnik\thanks{Oak Ridge National Labs} \quad}

\date{}			


\begin{document}

\maketitle

\thispagestyle{empty}

\begin{abstract}
We investigate and defend the possibility of causing a stock market crash via small manipulations of individual stock values that together realize an adversarial example to financial forecasting models, causing these models to make the self-fulfilling prediction of a crash. Such a crash triggered by an adversarial example would likely be hard to detect, since the model's predictions would be accurate and the interventions that would cause it are minor. This possibility is a major risk to financial stability and an opportunity for hostile actors to cause great economic damage to an adversary. This threat also exists against individual stocks and the corresponding valuation of individual companies. We outline how such an attack might proceed, what its theoretical basis is, how it can be directed towards a whole economy or an individual company, and how one might defend against it. We conclude that this threat is vastly underappreciated and requires urgent research on how to defend against it.
\end{abstract}

\bs
\bs
\bs

\newpage

\tableofcontents

\thispagestyle{empty}

\newpage 

\setcounter{page}{1}

\doublespace

\section{Introduction}
The deployment of advanced technologies such as machine learning (ML) based artificial intelligence (AI) in the financial industry has the potential to greatly benefit individual investors and the larger economy \cite{HSGAC2024_AI_HedgeFunds}. However, the accelerated use of these technologies, in particular relying on financial forecasting models for investment decisions, poses a significant and widely underappreciated risk to economic stability. We argue that such models are vulnerable to an attack first demonstrated in the domain of image classification, which would have significant negative consequences for the economy. In this paper, we outline how such an attack might be carried out, why we should think that it is possible, what its consequences would be, and how one might defend against it.

The well-being of a nation and its citizens is closely tied to the strength of its economy. The economy in turn is closely connected to the stock market. A booming stock market creates wealth, makes capital investments easier, increases tax revenue, and allows for increased spending on national security. An adversary can try to undermine this by manipulating the stock market, causing a sell-off, wealth destruction, and tighter national security budgets. Traditional attempts to manipulate the stock market include spreading false information, undermining confidence, spoofing, and a coordinated sell-off. In this paper, we outline a more technically sophisticated attempt to attack the stock market. It exploits the fact that present day stock market trading relies heavily on financial forecasting models, and the further fact that machine learning models are susceptible to adversarial examples. We develop how such an attack might proceed, why it is possible, and how one might defend against it. To carry it out fully would require further research, a dedicated and well-trained team, and possibly significant financial resources. However, these requirements can be met by large corporations or state actors, and so the possibility of this attack is a real but neglected danger to the economy of a nation. Because it can be directed not only at a whole market but also at a single stock, it is also a significant risk to individual companies since their market valuation is susceptible to such an attack from an adversary as well. Unfortunately, the risk of such an attack is currently not properly appreciated, and one of our main goals here is to make clear that this is a significant threat that needs to be taken very seriously.

In the following, we will outline how to carry out such an attack. Most of the technical aspects have been studied in the adversarial ML literature, and some have already been applied to the financial domain. Our goal is to connect these different parts to make vivid that there is a serious but neglected risk to individual companies and the whole economy. Although we will conclude that such attacks are very likely technically feasible, it is not clear how one can effectively defend against them. We conclude by outlining some defensive approaches, but more effective and properly developed methods are urgently needed from the perspective of both investors and regulators. Deferring to financial forecasting models in making investment decisions and the risk of adversarial attacks on such models is a special case of the larger problem that reliance on machine learning models in decision making can bring with it unintended risks that are not yet fully understood but that can be very significant.

\section{Background: Adversarial examples and financial forecasting models}

\subsection{Forecasting models and investment decisions}

Buying and selling decisions of investment firms are often based on the predictions of financial forecasting models. These models are machine learning models that aim to predict the price of a particular stock or of a stock index at some future time. There is a complex probabilistic connection between the price of an index or stock at a particular time and the price of the index and various other stocks at earlier times together with further facts. Machine learning models are well-suited to learn these complex connections and to make reasonably good predictions about these values because they rely on significantly larger data, require less human involvement, and operate much more rapidly on a larger scale and thus it is lucrative to invest on the basis of their predictions. Consequently, the use of financial forecasting machine learning models is widespread in the financial industry \cite{HSGAC2024_AI_HedgeFunds}. Models with various architectures such as long-short-term memory (LSTM), convolutional neural networks (CNN), and transformers are widely deployed to predict stock prices and other financial metrics \cite{reviewOfStockPrediction}.These models are trained largely on the basis of publicly available data: past stock prices, trading volume, general economic indicators such as GDP growth or federal interest rates, as well as other data such as topics in the news. A well-trained model will predict with a certain degree of confidence that the price of, say, the Nasdaq index or Apple's stock will go up or down at some particular future time, or it might more precisely predict what the value of the index or the stock will be at that future time. In light of these predictions, the model might recommend buying, selling, or holding that stock or investments in the index. Financial forecasting models have proven to be very useful investment tools and because of this, investing decisions are often made based on their recommendations. Investment decisions might not just be recommended by a forecasting model, but can be directly carried out by a connected agent model that is authorized to make transactions itself. This can be more efficient and even be the only option in high-frequency trading. This automation of stock market transactions carries with it an increased risk, which we will point to shortly. Although the details of the forecasting models deployed by large investment firms are generally proprietary and not known to outsiders, the fact that they are likely trained on public data is notable and will be important later on. 

If one could influence financial forecasting models to predict with high confidence that the stock market will move downward, then this would lead to strong recommendations to sell assets in the market or even a direct sale of them. If sufficiently many investors are influenced by these forecasting models, then this would lead to a significant sell-off. Such a sell-off in turn would depress the market, validating the predictions made by the models, and potentially leading to further sell-offs, depressing the market even more. If done on a sufficient scale, this could lead to major economic disruptions. Since reliance on machine learning models in investment decisions has become widespread, the possibility of gaining control or influence over the models poses a serious risk. The most obvious way to gain control over financial forecasting models is to breach cybersecurity barriers and directly influence their behavior. This means in essence hacking into the computing infrastructure of an investment firm or otherwise getting administrator access to them, which is a widely recognized threat. It should be extremely difficult to do this since this threat is clear and significant, extensively studied, and heavily defended against. We instead propose to consider a very different kind of attack. This attack does not rely on gaining control over the forecasting model directly through any kind of access breach. In fact, it assumes no direct access or even detailed knowledge of the architecture of the deployed forecasting models. Instead, our attack is based on a subtle influence on particular stock prices which will take control of the forecasting models in a way that is predictable even without having detailed knowledge of or direct access to them. The main idea goes back to the so-called {\em adversarial examples to machine learning models}.

\subsection{Adversarial machine learning}

Adversarial examples came to prominence around 2014 with the discovery that image classifiers can be systematically influenced to reclassify an image by adding noise to it that is virtually imperceptible to the human eye \cite{goodfellow2014generativeadversarialnetworks},  \cite{szegedy2014intriguingpropertiesneuralnetworks},  \cite{goodfellow2015explainingharnessingadversarialexamples}. An image classified with high confidence as a panda bear can be transformed into one that looks essentially the same to human observers but is now classified as a gibbon with high confidence. The method for this is to add noise to the former image, which is a small perturbation of the values of each pixel, either positive or negative. The values to be added are constructed from the image classifier and require access to the model in this construction. In the simplest case, we determine how changing the value of each pixel up or down contributes to the panda image being classified as a gibbon, and then add or subtract a fixed value in accordance to the value of that pixel. The resulting method for constructing an adversarial example --- the so-called {\em fast gradient sign method} (FGSM) --- is the most straightforward one for constructing such examples, and we will discuss it further below. Other, more sophisticated methods have been developed since then that optimize how an initial image should be modified to achieve a desired result. Such constructions correspond to adding what appears to be random noise and leads to a virtually indistinguishable change in the image but deeply affects the classification of that image. In general, we can distinguish between {\em untargeted attacks} and {\em targeted attacks}. An untargeted attack simply aims to fool the model and get it to classify an image of a dog, say, as an image of something else. A targeted attack aims to get the model to classify a particular image as belonging to some preselected class, e.g.~a goldfish. Amazingly, it has been shown in the literature that targeted adversarial examples can be constructed that work for any target class, regardless of what the original image is \cite{carlini2017towards}. We will go into more technical details below. 

Although adversarial examples for image classifiers have been widely discussed, and more recently language models \cite{zhang:aa}, they are a general feature of machine learning models, with some similarities and differences between different kinds of models. The simplest case of a financial forecasting model is a time-series model. Such a model is trained on sequences of values, say the price of the stock over a two-week interval, with the aim of predicting the price of the stock at a time after that sequence. The goal is to extend the sequence further into the future. To construct an adversarial example for such a model one would add noise, that is, a small perturbation, to the values of the stock in the time series, which in turn will lead the model to predict with some confidence that the stock will behave differently than it otherwise would. These small perturbations will be small values that are specifically constructed from the model in analogy with the noise added to an image. We need to determine whether increasing or decreasing the stock price at a particular moment contributes to getting the model to make the desired adversarial prediction. Such adversarial examples can again be either untargeted, aiming simply to mislead the model, or targeted, aiming to get the model to predict a particular desired outcome: either a particular target stock price or a strong sell or buy recommendation. We will discuss below in more detail why and how this is possible for financial forecasting models. In particular, we will discuss the construction of adversarial examples to forecasting models for single stocks and for indices, which are collectively determined by a whole group of stocks. 

In general, adversarial examples are specifically constructed data points that exploit the model's architecture to lead to behavior that is contrary to how it is intended to behave, and therefore adversarial. Adversarial examples are examples of data that are constructed and thus primarily synthetic data, not real-worldly data. This also applies to adversarial examples to financial forecasting models. In general such examples construct a time-series of financial data that has a certain desired adversarial effect on a forecasting model. But this time-series normally does not correspond to real financial data and does not correspond to actual stock prices or index values. But this does not have to be so, as it might well be that actual stock prices constitute an adversarial example to a particular forecasting model. We can say that an adversarial example is {\em realized} just in case the world is changed in such a way that the actual worldly data is an adversarial example to a forecasting model. An example is {\em unrealized} if it is simply synthetic data, corresponding to nothing real in the world. This distinction applies generally, including to adversarial examples to image classifiers. One prominent case of realized adversarial examples comes from facial recognition. In a well-known study \cite{sharif-accessorize}, researchers build glasses with pixelated color spots on them that adversarially affect facial recognition models. The pixels were placed in such a way that the model would falsely identify a face as belonging to actress Milla Jovovich and not to the completely different-looking person wearing the glasses. In this case, the glasses together with the wearer's actual face realize an adversarial example to a facial recognition model.

\begin{figure}
    \centering
    \includegraphics[width=0.6 \linewidth]{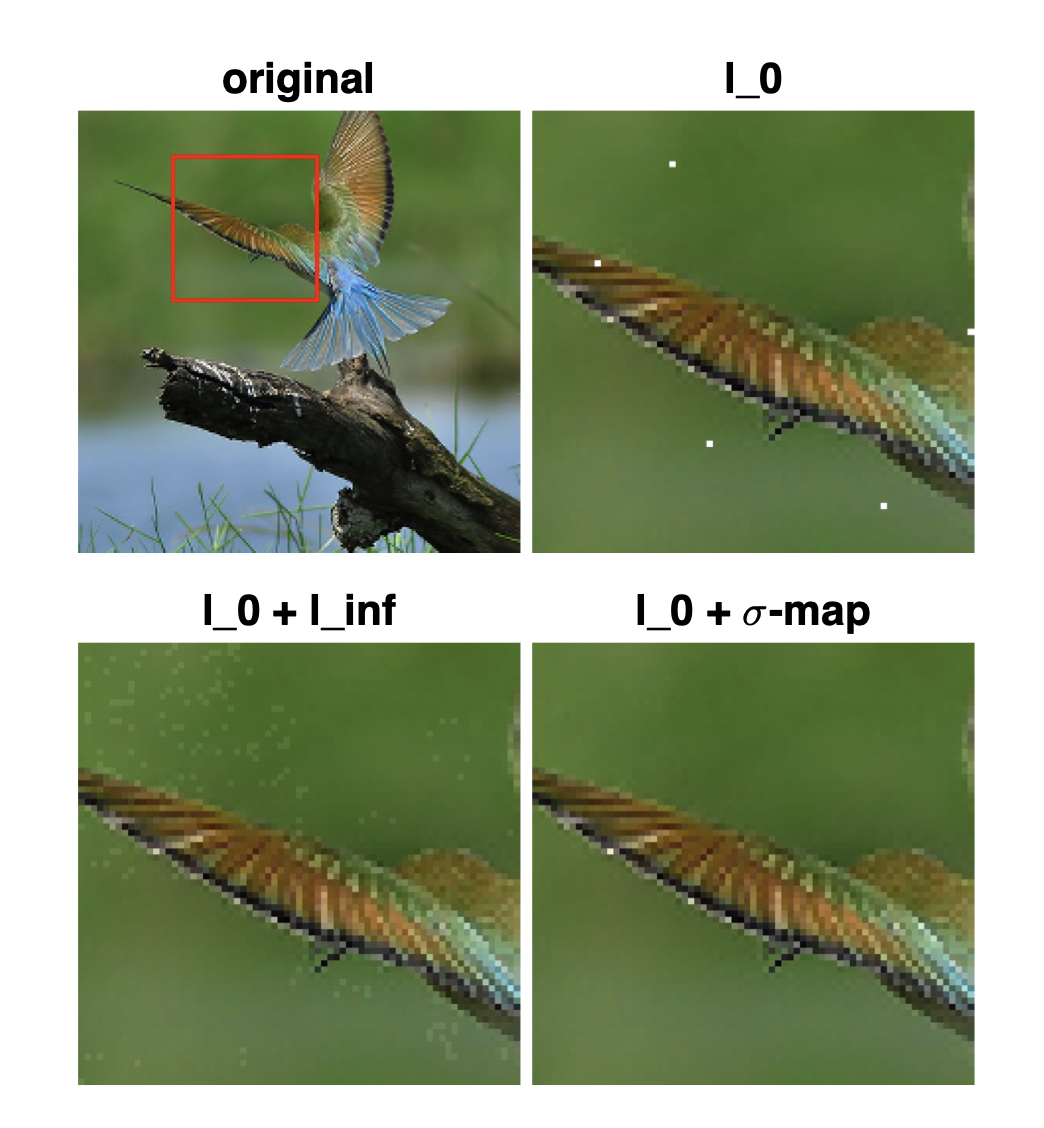}
    \caption{Three different sparse attacks: original, attacking very few pixels by a lot, attacking few by a bit, and attacking few by very little. The fourth, bottom right, image in particular highlights that the attack can be inperceptible and sparse. Image and research from \cite{Croce_2019_ICCV}. Image reused with permission.}
    \label{fig:sparse}
\end{figure}

This particular example also highlights a crucial general point about adversarial examples: one does not necessarily need to change every aspect of the input to produce the example. For an adversarial example to image classification it can be sufficient to modify part of the image, in extreme cases even just one pixel \cite{su2019one}. Similarly, in the adversarial glasses study, it is enough to add specifically color coded glasses to the face of a male researcher to be identified as a female actor, who on the face of it looks completely differently. Thus, modifying a small part of the image is sufficient to fool the classifier. We can say that an adversarial example is {\em sparse} if it is constructed from an original by modifying only a part of it and {\em very sparse} if it only modifies a small part of it. For image classifiers, it is known that there are very sparse adversarial examples \cite{Croce_2019_ICCV}. See Figure \ref{fig:sparse} for an illustration of what different sparse attacks look like for image classifiers, including the trade-off of how much one modifies a pixel and how many pixels one modifies for the attack to be successful. It should be surprising, even amazing, how little needs to be changed for such attacks to be successful. Just modifying a small fraction of pixels by an essentially imperceptible amount can be enough to completely change the classification of the image. This should not be possible, but has been soundly established in the literature, although its significance is not as widely appreciated as it should be. We will discuss a plausible explanation for why this works momentarily.

Although the construction of the adversarial examples outlined above requires knowledge of the details of the model that is being targeted and for which one aims to construct an adversarial example to, this is not in general required. It has been a surprising discovery that such examples are transferable to other models \cite{liu2016delving}. Although the panda-to-gibbon example was constructed using the details of the specific image classifier, it turns out that the same adversarial example also works for other models, even for other models that do not share the same internal architecture. All that seems to be crucial is that the two models were trained on similar data. This {\em transferability} of adversarial examples from one model to another is a key discovery in this area and crucial for us in the following. 

Adversarial examples can be seen as the result of a flaw in particular machine learning models, exposing some defective aspect of particular models. But this would be a mistake. Instead, adversarial examples should be seen as a natural consequence of machine learning. Adversarial examples are better understood as the result of the fact that in machine learning models can learn to recognize {\em non-robust features}: a correlation between a rather abstract but fragile feature of an image and its proper classification \cite{Ilyas2019_FeaturesNotBugs}. These non-robust features can be very susceptible to small perturbations. Although there is a strong correlation between having pixels with this particular distribution of values and being a picture of a panda, this correlation is fragile in the sense that small changes to these values break this correlation and are instead strongly correlated with being a gibbon picture. More robust features are less susceptible to such small perturbations, but in machine learning the model is normally not rewarded for picking out more robust features rather than less robust ones. Thus, the model will learn to recognize any feature, no matter how fragile, when learning classifications. The model is simply optimized to find correlations between features of an image and its classification, and without further care being taken, it will select robust or non-robust features, whichever work best for the task at hand. In particular, the features the model picks up on can be quite different from what our perceptual system is most sensitive too, and thus adversarial examples are classified very differently by human observers and machine learning models. We will return to this topic below. But we can see already that if the model picks up on non-robust features during its training, then it makes sense that an adversarial example can be constructed. Furthermore, it makes sense that such examples transfer to other models, even if these models have different architectures as long as they have similar training data. No matter what the internal architecture of the models is, both should pick up on the same features, robust or not, if they are strongly correlated with what the model is trained to predict in the training data. Transferability is thus no mystery but is connected to the explanation of why adversarial examples exist in the first place \cite{Ilyas2019_FeaturesNotBugs}.

\section{The basic idea of the Black Tuesday Attack}

We are now in a position to see how the Black Tuesday Attack could be carried out. A key insight of the Black Tuesday Attack is that one can realize an adversarial example to financial forecasting models by slightly manipulating the price of individual stocks. Instead of thinking of an adversarial example to financial forecasting models as a piece of synthetic data that is deployed to illustrate a vulnerability in a machine learning model, one can think of it as a realized state of the stock market, specifically brought about to mislead forecasting models. We argue that it is realistically possible for adversaries to realize such an example in the stock market even without having access to actually deployed financial forecasting models and thereby causing a crash of the stock market and significant economic damage, mirroring the actual Black Tuesday stock market crash of 1929 that was a key trigger for the Great Depression. All the key parts of such an attack are known to be possible. And although an attempt at such an attack cannot be guaranteed to work, there is likely little risk besides cost in trying to carry it out, and there is a good chance that it might work, although it is hard to say just how good that chance will be. With further research and some resources, one could potentially do a large amount of long-lasting damage to one's adversary, be it a nation or a company. We will outline the crucial parts of the attack now, without providing a detailed implementation. 

\subsection{Black Tuesday Attack Sequence}
The key steps for carrying out the Black Tuesday Attack are as follows: First, one needs to have access to a financial forecasting model. This does not require access to actually deployed models, i.e.~those in use by investment firms. Rather, one can simply train one oneself. All relevant data to train such a model are publicly available: history of stock prices, trade volume, economic indicators, etc. Training such a financial forecasting model is completely within reach with only limited hardware requirements and publicly available training data. Several financial forecasting models are also available in public repositories, although they may need to be modified or fine-tuned for the purpose of learning patterns in stock data \cite{reviewOfStockPrediction}. The model will be specifically trained for the particular forecasting task at hand. For our case, let us assume that the task is to predict the value of an index such as the Nasdaq. The Nasdaq is a weighted average of a large number of stocks and other values, weighted by market capitalization. Larger companies have a greater impact on the final value of the Nasdaq, but smaller companies might nonetheless be predictively highly relevant. The value of the Nasdaq is a great indicator of overall economic performance. A steep dip in Nasdaq value is a clear sign of an economic downturn, and advance knowledge of it is a good reason to get out of the market and sell one's assets of stocks. The target of the version of the Black Tuesday Attack that focuses on the Nasdaq is thus to cause financial forecasting models to predict a steep decline of the Nasdaq. The method to achieve this is to realize an adversarial example to these forecasting models in actual stock values, which leads the model to predict with high confidence that the index will decline steeply. 

The second step is to get a clear sense of how much it would cost to manipulate the stock price of various stocks by a given $\epsilon$, either up or down. The goal of such a stock manipulation is not to make money. In fact, the opposite should be expected, and prior work on adversarial attacks for trading already incorporates explicit cost constraints and "capital required" when crafting perturbations \cite{Goldblum_2021}. It would be costly to change the price of a stock by a given amount, leaving aside any considerations about whether this is a good investment. Manipulating stock prices simply to change their value, without the goal of financial gain, is illegal in the US. But, of course, this is only a small obstacle in our scenario, where adversarial actors aim to harm the economy. The legal restrictions to do so should have little effect. The cost of changing a given stock by a fixed small amount will vary widely from stock to stock, with smaller, less heavily traded stocks generally being much easier to influence than larger ones. Ideally, one would have two classes of stocks: those that it is feasible to manipulate sufficiently and those whose manipulation is out of reach given the resources one has available. Let us call the former group the {\em manipulable stocks} and the latter the {\em non-manipulable stocks}. 

The third step is to construct an adversarial example to the Nasdaq forecasting model that changes only the values of manipulable stocks. Such an attack will be both sparse and targeted. It is targeted since it does not just aim to mislead the model but to mislead it in a particular way: to predict a sharp decline. It is sparse, since it only changes some of the manipulable stock values, ideally as few as possible. Such a sparse and targeted adversarial example to forecasting models is analogous to the example of adding specifically designed glasses to a face to fool facial recognition models into identifying a particular predetermined person. Here, one knows in advance that one can only change the pixels around one's eyes, via the glasses, but one can still construct examples that are sufficient for an erroneous targeted classification. Similarly in our case, where one needs to construct an adversarial example changing only the values of manipulable stocks, leaving the values of the other ones alone. To construct such an example one can rely on the white box access to the financial forecasting model one has obtained in step one. Using the gradient of that model and the restriction to change only manipulable stocks, one can see how much and in what way these stocks need to be manipulated to get the model to predict with confidence that the index will decline in the future. Different methods for constructing such examples will be discussed in more detail below. The less change of the stock price of particular stocks is required, the more feasible the manipulation will be. Thus, the target will be to keep $\epsilon$ small. It can be that the size of $\epsilon$ and the distinction between manipulable and non-manipulable stocks need to be determined together and depends on one's budget. It is known from other cases of adversarial examples that small modifications to a small number of data points can be enough to fool the models. Moreover, previous work has demonstrated perturbations targeted against algorithmic traders on S\&P 500 constituents, indicating feasibility for attacks that generalize to target indices \cite{nehemya2021takingoverthemarket}.

Finally, one needs to have a coordinated intervention in the stock market to realize the adversarial example. This requires intervening by buying or selling the proper amount of the right stocks at the right time to realize the adversarial example in the actual market. Here, one needs to carefully consider how one's own intervention in the market will affect the value of a stock to achieve the target price. This has been investigated in the literature; see the differentiable stock market model in \cite{Goldblum_2021}, and although there is quite a bit of uncertainty as to how other independent interventions in the market will affect achieving one's target value, this can be mitigated simply by trying repeatedly if the original target value range is not being achieved. Depending on the adversarial example constructed, such a realization could all be coordinated at a single time or require planning over a series of times. It is possible that a sparse intervention of several manipulable stocks at the same time is sufficient to realize the example, and if so, then this should be doable. One will need a coordinated attempt to change the values of these stocks by the required small amount, relying on information about outstanding buy and sell orders and how one's intervention at a particular time will affect the value of the target stock. It might not be trivial to do this, but it should be doable \cite{Goldblum_2021}. 

\begin{figure}
    \centering
    \includegraphics[width=0.9\linewidth]{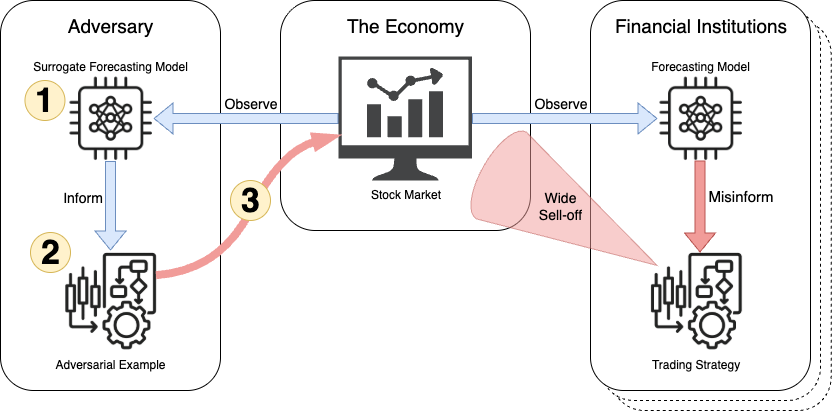}
    \caption{Black Tuesday Attack flow:
    1) Train a surrogate forecasting model; 
    2) Construct adversarial example; 
    3) Realize adversarial example in Stock Market}
    \label{fig:bta-outline}
\end{figure}

\begin{table*}
\centering
\small
\begin{threeparttable}
\caption{Black Tuesday Attack (BTA): Threat Model}
\label{tab:threat-model}
\begin{tabularx}{\textwidth}{@{}l X@{}}
\toprule
Component & Specification \\
\midrule
Adversary & A well-resourced actor such as a nation state, large company, or organization. \\
\addlinespace[2pt]
Goals & Destabilizing the stock market by causing financial forecasting models to confidently predict an impending sell-off.\\
\addlinespace[2pt]
Target &  Financial forecasting models used by financial institutions to aid in financial decision-making.
\\
\addlinespace[2pt]
Method & Craft targeted, sparse adversarial perturbations with a surrogate financial forecasting model, and execute trades that realize those perturbations in the real market. Perturbations dynamically adapt to evolving market conditions.\\
\addlinespace[2pt]
Knowledge & Publicly available market data. No insight into the models used in financial decision-making\\
\addlinespace[2pt]
Attack surface & Adversarial perturbations realized in the market become model inputs to forecasting models at financial institutions.\\
\addlinespace[2pt]
Constraints & Limited capital, adversarial robustness of forecasting models.\\
\addlinespace[2pt]
Success metric & (I) Some fraction of market participants shift strategy to sell. (II) Target index experiences a certain percentage drop in a short horizon.\\
\bottomrule
\end{tabularx}
\end{threeparttable}
\end{table*}

\subsection{Attack Discussion}
If the adversarial example that one aims to realize is diachronic, then this requires planning and a series of interventions at different times. Since one can only intervene at the present time, one needs to carefully plan a series of interventions that modify present values and over time produce a modified time series which realizes an adversarial example. Here, there will be further obstacles that can affect one's chances of success, but these, too, can likely be overcome. One extra difficulty affects the overall setup of the attack. It is possible that during the time one carries out the different steps of the attack, aiming to modify a select group of stock values at different times, one after the other, that some stocks change dramatically in value and trading volume so that they no longer are manipulable with the resources one has available to achieve the target value. The distinction between manipulable and non-manipulable stocks was made based on one's resources for achieving a desired change in price, and that itself can change over time. If some stocks become out of reach for manipulation during the diachronic attack, then the attack might fail. 

This highlights that it will be difficult to set up an attack that is guaranteed to succeed, but such a guarantee is not needed. The risk to economic stability is that such an attack might succeed. It is sufficient that it could be carried out under favorable circumstances. Since any attempted interventions to realize an adversarial example are likely very hard to detect, it is perfectly possible to try repeatedly until one actually succeeds. Such success might depend on favorable circumstances, say that no one else intervenes substantially with one of the target stocks to undo one's own attempts to reach a certain value of that stock. Here, it might well be wise to carry out the attack at a time when extra difficulties such as very active trading are minimized. 

A second difficulty of a diachronic attack affects the influence that other stocks, which one is not aiming to manipulate, have on the  adversarial example one aims to realize. To illustrate, if we are attacking the forecasting of the value of an index, then this forecasting model will take into consideration the values of many stocks, most of which we will not touch during the process of realizing an adversarial example to the model. Here, too, there can be obstacles that ruin one's chances of success, but success does not have to be guaranteed for the risk to be real. To illustrate the difficulty, one might plan to intervene in the next 10 time-steps of the values of 10 selected stocks. Those intended modifications would constitute an adversarial example to an index forecasting model assuming that the values of the other stocks stay within a certain range. However, after 5 successful interventions, one of the other stocks moves outside of that range. The originally planned adversarial example might then not work any longer after time-step 6, but it might well be possible to adopt ones plan in light of the new range of the stocks and modify the 10 target stocks in a slightly different way after time-step 6 to still achieve an adversarial example overall after 10 time-steps. This dynamic nature of the example to be realized can overcome the obstacles that a fixed planned adversarial example will face. Such a dynamic approach can start with a planned example, and then modify the plan at each time-step as need be, in light of other changes in the stock market. All one needs is a successful path, starting at time-step one, and then extending it one step at a time, leading to a successful attack overall. Since there will be not just one adversarial example, but many, this dynamic strategy should be seen as promising. To our knowledge, this has not been adequately studied in the literature to date. 

Another approach to a successful diachronic adversarial example to forecasting models is to aim for a universal adversarial example on where certain sparse modifications work no matter what the values of the other data points are. If such an example can be constructed modifying only a small number of stock values, then it would fool the forecasting model no matter what the values of the other stocks turn out to be during the time the attack proceeds. Such universal adversarial examples to time-series forecasting models are discussed in \cite{nehemya2021takingoverthemarket}, but they are likely harder to realize and less effective than a dynamically adjusting example. Again, to our knowledge this has not been properly studied yet. 

After the adversarial example has been realized in the market, it should affect not just the financial forecasting model that was used to construct it, i.e.~the model that is white box accessible to the adversary, but it should also be transferable to forecasting models that are not directly accessible and thus black boxes to the adversary. The transferability of adversarial examples has been widely studied (\cite{carlini2017towards}, \cite{liu2016delving}) and shown to apply to time-series forecasting models as well as to image classifiers. Since these models are likely trained on the same or similar training data, namely the public data of past stock values and other economic indicators, adversarial examples should be very transferable across models, including across model architectures. That would mean that other financial forecasting models should be affected by the example and thus confidently forecast that the Nasdaq will fall. 

Such a prediction of the forecasting models will then with some likelihood lead to a major market depression throughout. Since the Black Tuesday Attack focuses on a major stock index, not a single stock or a group of stocks, the prediction of the steep decline of the index should trigger a sell-off across the market, not just a localized one. And since financial forecasting models can be authorized to engage in trading directly, without direct human intervention \cite{HSGAC2024_AI_HedgeFunds}, the sell-off might be triggered automatically, further exposing the risk of reliance on machine learning models \cite{HSGAC2024_AI_HedgeFunds}. An initial decline of the index due to such triggered sell-offs will verify the model's prediction and justify further sell-offs. 

It is hard to predict what stopgaps might be effective in minimizing the damage and if a drop would be short-lived or longer lasting. It might seem that it should be only a momentary dip, since once it is clear that the dip is due to an adversarial examples, a buying spree will undo the losses. However, it will be hard to detect that the dip is due to an adversarial example. First, the model's predictions will become true, and so it is hard to put blame on the models themselves and see evidence that they have been fooled. Second, it would be hard to detect the market manipulation of individual stocks since they can each be manipulated separately in a coordinated way. Third, as of now, even the possibility of such an attack is hardly on anyone's radar. If the attack goes undetected, then it might well lead to the sentiment in the market that the drop of the index is justified, and thus it will not be corrected. How long a dip would last and whether it would be undone by a future buying spree cannot be predicted. The Black Tuesday Attack could only lead to a momentary loss of value, or it could trigger a recession and lasting damage. To test the attack, one might aim first for a smaller market, say an index with fewer stocks and lower market capitalization, or a smaller local economy, before attacking a major index like the Nasdaq. This would not only allow one to test the success of the attack itself, but also the longer consequences. 

With this outline, we hope to have made the case that the Black Tuesday Attack is a realistic possibility. See Figure \ref{fig:bta-outline} for an illustration of how this attack works and Table \ref{tab:threat-model} for a characterization of the threat model. Many of the key parts of it have been studied and in principle verified. Some parts of it are understudied, but do not seem to face in principle obstacles. If the attack succeeds, it could cause tremendous economic damage and thus is a real risk that urgently needs to be addressed and defended against. Although carrying out the attack in practice will require more research and likely substantial resources, it is nevertheless likely easier to carry out than other attacks that cause comparable damage.

\subsection{Single stock attack}

Although our focus above was on attacking an index and a whole economy, the threat can easily be modified to the less ambitious goal of attacking the valuation of a single company and their stock. The threat of devaluing a stock with traditional methods is well known and significant: ``A successful massive misinformation scheme for a widely held company like Apple, Facebook, or General Electric could have a monetary impact measured in the billions of dollars and affect a significant population of investors since those companies make up large positions in retirement accounts. Furthermore, unlike many other methods of market manipulation, mass misinformation can be motivated by goals of personal profit as well as goals of non-profit disruption'' \cite{lin2016new}. 
Such an attack would be easier to carry out since one only hopes to affect the prediction of a single stock value, and it would also be more likely to occur since there are numerous possible incentives other than simply causing damage to do so. First, it might be an effective way towards financial gain: one could short a particular stock, realize an adversarial example to cause forecasting models to predict its decline, and thereby trigger a sell-off of that stock and its decline. Whether this is a viable strategy for profit will depend on whether it is cost-effective to realize the adversarial example and profit from the short selling afterwards. This is hard to predict in advance, since it is not clear how much manipulation would be required to realize the adversarial example and how costly it would be. Second, an economic adversary of one company might benefit from a drop in valuation of a rival company and thereby be motivated to attack the valuation of this rival company. To successfully carry out this attack would not require a net profit in the stock manipulation but might make sense simply because it would weaken one's competitor in the marketplace. Obviously, the legality of these attacks is highly dubious, but this is not our topic here. We simply hope to find out if such attacks are feasible at all and who might be motivated and able to carry them out. For the case of attacking a single stock the damage overall will be more limited, but the likelihood of someone carrying it out might well be noticeably higher since we predict it to be easier and there will be more actors able and willing to carry them out. 

The mechanism of a single stock attack is exactly parallel to the Black Tuesday Attack: train your own forecasting model on a similar dataset than you expect investment firms to train theirs, then use the white-box access to your own model to construct a sparse adversarial example to this model. This example should be transferable to the models actually deployed. Finally, realize the example in the market via a controlled manipulation of the few stocks that need to be modified to achieve this example.

Single-stock manipulations are a security threat to every company whose stock is publicly traded \cite{lin2016new}. Its own valuation can be substantially impacted by the result of such an attack. However, companies generally do not have control of the financial forecasting models that predict their stock price. Those models belong to investment firms that use them for investment decisions. This gives rise to an intriguing issue regarding how to defend against such attacks.

\section{Defending against the attack}

We hoped to make the case above that one should seriously consider the possibility of the Black Tuesday Attack as a threat to the economy. Such an attack should be considered possible, and if successful, it could have potentially catastrophic consequences to the economy and the well-being of societies that depend on it. We also hoped to make the case that a similar attack can be carried out against the stock of individual companies, greatly endangering their valuation and place in the market. Thus, it is of utmost importance to prepare and defend against it. In this section, we discuss several ways in which one might do this. We will discuss several techniques to mitigate the risk of adversarial attacks on machine learning models in general, which are discussed in the literature on adversarial examples and whether and how they might be applicable in the context of financial forecasting models. As will become clear, none of the known defense methods provide a reassuring defense against such an attack.

The most direct defense against adversarial attacks is {\em adversarial training}: include adversarial examples in the training data. The goal of adversarial training is to force the model to pick up on more robust features to make its prediction. As we discussed above, adversarial examples are possible, since models learn to pick up on any predictively relevant feature in their training, no matter how fragile. If the training data itself contain adversarial examples, this means that certain non-robust features are being excluded from being predictively successful, and so the model needs to find more robust features, ones that are not fragile to small perturbations. To do this for financial forecasting models, one would need to add adversarial examples, ones specifically constructed as synthetic data, to the worldly data of actual stock values, which constitute the training data for the forecasting model. This naturally gives rise to the concern that performance will be diminished, since the new model is trained on essentially made-up financial data. Since the Black Tuesday Attack is presumably unlikely to happen, it is hard to see that performance degradation will be acceptable in practice. 

A second option of defense is simply detection. One could be on the lookout for market manipulation that aims to realize an adversarial example in the market. But this is likely not easy to do since the manipulation does not have to be carried out by a single actor. A group of investors can coordinate to manipulate a collection of stocks in a predetermined way, spreading their investment activities in possibly a complex coordinated way. It might still be possible to have some evidence of such manipulation. For example, a coordinated manipulation might quickly buy or sell a stock to reach a particular target value, and it might be detectable that such a price target is intended by a series of trades. Although such detection cannot be ruled out, it appears to be challenging. 

A third option is to stick with worldly data in model training, but to smooth out values when predictions are made. This would involve a controlled small modification of the actual stock values on which predictions are made with the idea that such smoothing will destroy the possible adversarial example. Although this in general is a valid technique \cite{liu2023robustmultivariatetimeseriesforecasting} to bypass adversarial examples, it is more easily applicable in image classification where smoothing of an image has little or no effect on the perceived quality of the image than it is applicable to financial forecasting, where small changes could be predictively significant. One should again expect performance degradation of the forecasting model, since that model is likely to be predictively the most successful when predicting from raw data, not smoothed data.

Overall, more research is needed to get a clearer sense of what realistic defenses against adversarial attacks to financial forecasting models there are that do not lead to performance degradation and whether forecasting models can be built that are less susceptible to adversarial attacks. The resulting situation will likely mirror the present state of research on adversarial examples, where successful defenses against one kind of attack are being developed only to lead to new kinds of attacks which require new kinds of defenses. In our discussion above, we mentioned constructing adversarial examples using the simplest established technique: FGSM. Adversarial training against such examples is a successful defense, but it is not necessarily successful against other more sophisticated methods to construct new adversarial examples \cite{madry2018towards}. This leads to an arms race of new attacks and new defenses in financial forecasting as well as in other domains where machine learning models are deployed.

\section{Conclusion}

Adversarial attacks on financial forecasting models are a serious and substantial threat to economic stability, a threat that is not well known and appreciated and also, at present, not defended against. We have outlined the main steps for carrying out the Black Tuesday Attack, a sophisticated attack on financial forecasting models that could potentially lead to a stock market crash. We tried to make the case that all the crucial steps for such an attack (sparseness, transferability, ability to construct the adversarial examples in the first place, etc.) are essentially known to work in principle for a large range of machine learning models, including financial forecasting models.  Although these attacks have been studied widely for image classification models, they apply to financial forecasting models just as well, as recent literature has shown. There is a clear route to how one might realize an adversarial example in the actual stock market. Transferability makes clear that one can construct adversarial examples to forecasting models even without having white-box access to ones that are actually deployed. We believe that this general threat is largely neglected in financial and national security but deserves much greater attention. Although we have only outlined the attack at a basic proof-of-concept level, a more detailed implementation is likely possible in the near future. We urgently need more research on effective defenses against such an attack and a better understanding of its potentials and limitations.

\addcontentsline{toc}{section}{Bibliography}

\bibliographystyle{apalike}

\bibliography{biblio2}

\end{document}